\documentclass[twocolumn]{aastex631}

\usepackage{enumitem}

\shortauthors{Mercier et al.}
\graphicspath{{./}{figures/}}

\begin{document}

\title{Revisiting the Iconic Spitzer Phase Curve of 55 Cancri e: Hotter Dayside, Cooler Nightside and Smaller Phase Offset}

\author[0000-0002-0962-7585]{Samson J. Mercier}
\affiliation{Department of Physics, McGill University, 3600 rue University, Montréal, QC H3A 2T8, Canada}

\author[0000-0003-4987-6591]{Lisa Dang}
\affiliation{Department of Physics, McGill University, 3600 rue University, Montréal, QC H3A 2T8, Canada}

\author{Alexander Gass}
\affiliation{Department of Physics, McGill University, 3600 rue University, Montréal, QC H3A 2T8, Canada}

\author[0000-0001-6129-5699]{Nicolas B.\ Cowan}
\affiliation{Department of Physics, McGill University, 3600 rue University, Montréal, QC H3A 2T8, Canada}
\affiliation{Department of Earth \& Planetary Sciences, McGill University, 3450 rue University, Montréal, QC H3A 0E8, Canada}

\author[0000-0003-4177-2149]{Taylor J.\ Bell}
\affiliation{Department of Physics, McGill University, 3600 rue University, Montréal, QC H3A 2T8, Canada}
\affiliation{BAER Institute, NASA Ames Research Center, Moffett Field, CA 94035, USA}

\begin{abstract}
Thermal phase curves of short period exoplanets provide the best constraints on the atmospheric dynamics and heat transport in their atmospheres. The published Spitzer Space Telescope phase curve of 55 Cancri e, an ultra-short period super-Earth, exhibits a large phase offset suggesting significant eastward heat recirculation, unexpected on such a hot planet \citep{Demory2016b}. We present our re-reduction and analysis of these iconic observations using the open source and modular Spitzer Phase Curve Analysis (SPCA) pipeline.  In particular, we attempt to reproduce the published analysis using the same instrument detrending scheme as the original authors. We retrieve the dayside temperature ($T_{\rm day} = 3771^{+669}_{-520}$ K), nightside temperature ($T_{\rm night} < 1649$ K at $2\sigma$), and longitudinal offset of the planet’s hot spot and quantify how they depend on the reduction and detrending. Our re-analysis suggests that 55 Cancri e has a negligible hot spot offset of $-12^{+21}_{-18}$ degrees east. The small phase offset and cool nightside are consistent with the poor heat transport expected on ultra-short period planets. The high dayside 4.5~$\micron$ brightness temperature is qualitatively consistent with SiO emission from an inverted rock vapour atmosphere. %Check abstract word limit - if have room can add the below - From our orbital phase curve, we are also able to constrain the eclipse depth and phase semi-amplitude to $209^{+50}_{-47}$ p.p.m. and $110.9^{+17.2}_{-15.6}$ p.p.m., respectively.

\end{abstract}



%%%%%%%%%%%%%%%%%%%%%%%%%%%%%%%%%%%%%%%%%%%%%%%%%%

%%%%%%%%%%%%%%%%% BODY OF PAPER %%%%%%%%%%%%%%%%%%

\section{Introduction}\label{Intro}
% \begin{itemize}
%     \item \sout{Paragraph on transiting exoplanets: small planeta are hard, but small hot planets are sometimes feasible.}
%     \item \sout{Discovery of 55 Cancri e: RV, aliasing: e was USP!  transits!}
%     \item \sout{Summarize debate about radius and mass (hence density and bulk composition) of 55 Cnc e: Crida+2018 estimate envelope mass via Bayesian voodoo; Bourrier+2018 update mass and radius and hence bulk density; Morris+2021 seems to be latest/greatest radius measurement.}
%     \item \sout{Summarize atmospheric characterization of 55 Cnc e: Ehrenreich+2012 reported non-detection of Ly$\alpha$; Tsiaras+2016 reported detection of HCN in Hubble transit spectrum; Ridden-Harper+2016 report non-detection of Na and Ca in transit spectra; Esteves+2017 report non detection of water in transit spectrum; Sulis, Dragomir et al. 2019 report MOST observations, including seasonal variations of star-spots (?); Tabernero+2020 upper limit on Na; Zhang+2021 do not detect escaping He; Deibert+2021 obtain upper limits on lots of molecules, including HCN.} 
%     \item \sout{Segue to Demory phase curve paper and key results; also mention re-analysis by Angelo \& Hu (2017)}
%     \item \sout{Explain why 55 Cnc e is notoriously hard phase curve to reduce/analyze}
%     \item \sout{Outline our paper (this may not be required for such a short paper, or could be folded into the previous paragraph)}
% \end{itemize}

%Possible referees : Heather Knutson, Ian Crossfield, Renyu Hu, Vincent Bourrier

Rocky ultra-short period planets (USPs) are the most amenable terrestrial exoplanets for atmospheric study (for a review, see \citealt{Winn2018}). Full-orbit Spitzer phasecurves of USPs LHS 3844b \citep{Kreidberg2019} and K2-141b \citep{Zieba2022} exhibit no significant hot spot offset and therefore suggest poor day-to-night heat transport. In contrast, 4.5~$\micron$ \textit{Spitzer} observations of the best-characterized USP planet, 55 Cancri e \citep{Fischer2008, DawsonFabrycky2010, Winn2011, Demory2011}, revealed an eastward phase offset of $\theta_{\rm phase} = 41 \pm 12 $ degrees \citep[][abbreviated as D16b hereafter]{Demory2016b}. %A few years into the \textit{Warm} Spitzer mission, \cite{Demory2016b} reported the thermal map of 55 Cnc e which suggest noticeable atmospheric heat transport. In contrast, 

 The Spitzer data, as reduced by D16b, were also re-analyzed by \cite{AngeloHu2017} using the phase curve parameterization of \cite{Hu2015}. They found an average dayside temperature and a phase curve offset in agreement with D16b, but they obtained a smaller phase curve amplitude suggesting a hotter nightside temperature: $1610^{+120}_{-130}$ K \citep{AngeloHu2017} cf.\ $1380^{+340}_{-450}$ K (D16b). If the Spitzer phase modulation is due to the planet, it suggests effective atmospheric heat circulation. This would be surprising: hot rocky planets are expected to have thin atmospheres  \citep{castan2011atmospheres} and short dayside radiative timescales  \citep[][]{Cowan2011}. 

\begin{figure*}
    \centering
    \includegraphics[width =\textwidth]{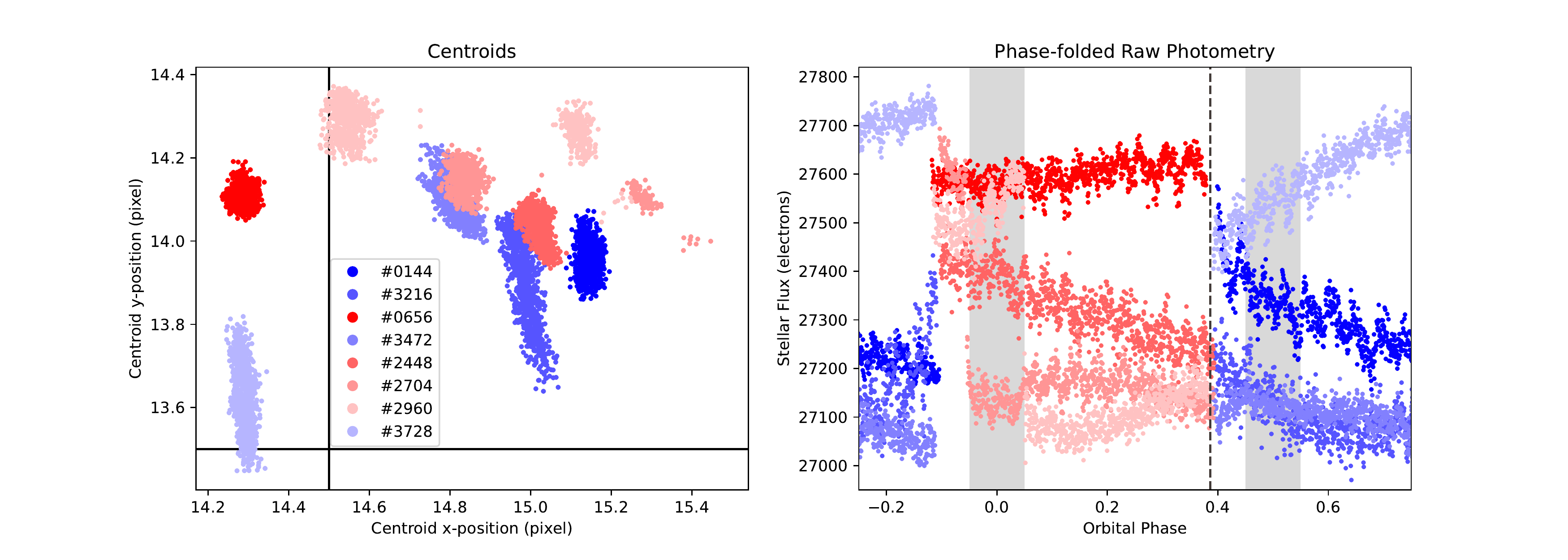}
    \caption{\textbf{Centroid Position and Raw Photometric data | }Plotted on the left is the centroid position on the IRAC detector and on the right is the phase-folded raw photometry obtained from SPCA. In both plots, the 8 AORs are separated into two groups: the red group represents the phase range -0.1 to 0.4 and the blue group represents the phase range 0.4 to 0.9. The solid black lines on the left plot show the pixel edges. The grey bands in the right plot correspond to the transit (0 phase) and eclipse (0.5 phase) of 55 Cancri e. The black dashed vertical line represents the phase offset obtained by D16b. Note that the offset lines up with the suture point of the two AOR groups around 0.4 orbital phase.}
    \label{fig:key1and2}
\end{figure*}

%\textcolor{cyan}{paragraph 4: why we think the Spitzer Phase Curve is sus. Start with explaining the challenges of removing the infamous spitzer detector systematics.}\\
While the phase variations of 55 Cnc e raise many questions, Spitzer IRAC observations are known to be plagued with detector systematics due to the interplay of telescope pointing fluctuations with intra-pixel gain variations. Extracting exoplanets signals at the level of 100 parts per million (ppm) can be therefore extremely challenging and requires the removal of significant instrumental effects (for a review, see \citealt{Ingalls2016}). Over the past decade, improvements in observing strategy and development of a suite of techniques to remove IRAC time-correlated noise have yielded robust thermal phase curves for many short-period exoplanets \citep{May2022}. However, in addition to the challenges of observing small planets, the 55 Cnc e Spitzer data are particularly complicated to analyze:
\begin{enumerate}
    \item Since 55 Cnc is a very bright star, the exposure time was reduced to 0.02 s to avoid saturation. Shorter exposures meant more images and hence frequent interruptions of the observations to down-link data to Earth due to Spitzer's limited capacity. This, in turn, resulted in the telescope performing multiple visits of 55 Cnc e over the course of a month, as shown in Figure \ref{fig:key1and2}. Each visit corresponds to an Astronomical Observation Request (AOR).
    %Exposure times for Spitzer phase curve measurements are typically 2 s. However,  
    \item Because the data were collected over multiple AORs, the target's position on the IRAC detector varied between observations. The increased spread in centroids means that one has to model the sensitivity of multiple non-overlapping areas of the detector, shown in Figure \ref{fig:key1and2}. Since different parts of the detector behave differently, this increased centroid position spread adds significant uncertainty to our planetary inferences.
\end{enumerate}
In this paper, we present a re-reduction and re-analysis of the Spitzer/IRAC phase curve of 55 Cnc e published in D16b but using the Spitzer Phase Curve Analysis pipeline \citep[SPCA\footnote{\url{https://spca.readthedocs.io/en/v0.3/}}:][]{dang2018, belldang2020}. We first outline our data retrieval, reduction and analysis methods in Section \ref{M&R}. In Section \ref{D&C} we present our results and discuss the implication of our retrieved phase curve. %why our retrieved parameters differ from previous studies as well as our conclusions regarding 55 Cnc e's bulk composition and atmospheric composition.

%\begin{itemize}
%    \item Explain how the data were retrieved (NASA Exoplanet Archive, Spitzer). Mention which version if Spitzer data calibration pipeline was used... it might matter for these data!
%    \item Briefly explain our data reduction scheme - reference SPCA papers
%    \item Explain briefly how we chose our photometric reduction scheme for all the AORs
%    \item Explain what model we used to de-trend our data and the free parameters, fixed parameters and hyper-parameters used for said model.
   
%    \item Describe key plots and main table
%\end{itemize}

\section{Methods}\label{M&R}
\subsection{Data Retrieval}\label{Retrieve}
We used observations of the 55 Cnc system (PID 90208, \citealt{Demory2012prop}) originally presented by D16b and collected with the 4.5~$\micron$ channel of the InfraRed Array Camera \citep{Fazio2004} aboard the Spitzer Space Telescope \citep{Werner2004}. The observations were taken from 15 June to 15 July 2013 and consist of 8 AORs of continuous time-series observations, each approximately 9 hours in duration, corresponding to half the duration of 55 Cnc e's orbital period, and with a gap of 3 to 8 days between AORs. The basic Spitzer data calibrated with the S18.18.0 pipeline version were retrieved from the Spitzer Heritage Archive\footnote{\url{https://sha.ipac.caltech.edu/applications/Spitzer/SHA/}} \citep[DOI:][]{Spitzer_doi}. The observations were taken in subarray mode with an exposure time of 0.02 s for a total of 4,918,760 frames. We note that AORs 48072704 and 48072960 experienced 30-min long interruptions during observations and therefore exhibit an important jump in the centroids (see Figure \ref{fig:key1and2}). 

\subsection{Data Reduction}\label{Reduce}
%The basic calibrated data (BCD) extracted from the Spitzer Heritage Archive are a Flexible Image Transfer System(FITS) formatted into datacubes. The datacubes contain 64 images, each 32-by-32 pixels. We reduced our data using the SPCA pipeline. 

Using SPCA, the Spitzer .fits files are first unzipped and loaded into RAM. An initial $4 \sigma$ clipping is done along the time axis for each AOR to mask artifacts such as cosmic ray hits or hot pixels. If a masked pixel is within a $5\times5$ pixel square centered on the target, the frame is ignored entirely. We also evaluate the flux-weighted-mean centroid of the target and the point response function (PRF) width along the $x-$ and $y-$axis for each frame. Next, frames with flux, centroids or PRF widths differing from the datacube median value along the time axis by $4\sigma$ are flagged. We perform photometric extraction with and without these flagged frames, and ultimately choose the photometric scheme resulting in the lowest scatter as explained in the following subsection. Finally, the background flux is subtracted frame by frame and a last $5\sigma$ sigma clipping is performed to remove any remaining artifacts. %SPCA then performs photometry and the data is binned by datacube(factor 64 binning). 
A more in-depth description of the reduction process is provided in \cite{dang2018} and \cite{belldang2020}. The data from all eight AORs totals approximately 70 GB. In order to diminish the computational cost of the model fitting and detrending phase, we chose to bin the data by a factor of 640, or 12.8 s bins (further detailed in Appendix \ref{sec:appendix}). 

\subsection{Photometry}\label{Photo}
%As SPCA allows us to choose from three photometric reduction schemes: Aperture photometry, Point-Spread Function(PSF) photometry
We experimented with Point-Spread Function (PSF) fitting photometry and a variety of circular apertures (soft-edge and hard-edge) with radii ranging from 2.0 to 6.0 pixels. We evaluate the RMS scatter in flux for each photometric scheme of each AOR. We also experiment with a fixed aperture centered on the (14, 15) pixel and a moving aperture centered on the respective centroid of each frame. Ultimately, we find that a moving, soft-edge aperture of radius 2.2 pixels gave us the lowest scatter averaged over all AORs. We therefore adopt this photometric scheme, shown in the right plot of Figure \ref{fig:key1and2}, for the remainder of our analysis.

We fit the reduced data with a two-component model accounting for both detector and astrophysical variations. The detector component is essential to remove the intra-pixel sensitivity variations in Spitzer/IRAC data. As shown in the left plot of Figure \ref{fig:key1and2}, the centroids are dispersed and 2 AORs are separated into multiple regions (48072704 and 48072960). The large spread of centroids underlines the importance of detrending the detector systematics from the photometric data to distinguish an astrophysical signal. In the right panel of Figure \ref{fig:key1and2}, we highlight the two groups of AORs with no overlap in phase: those spanning orbital phases of -0.1 to 0.4, or 0.4 to 0.9. The phase offset obtained by D16b, shown by the black dashed vertical line, occurs at the suture phase for the two groups of AORs, which is suspicious and reinforces the need for a re-reduction and re-analysis.
% and Pixel-Level Decorrelation(PLD) photometry.
%These photometric methods allow us to extract flux values from each datacube. However, due to the large size of the 55 Cancri e data set we had to treat each AOR separately, meaning we had to choose which photometric scheme to consistently use for all of them. 
%To do so, we selected the method that, on average, gave the lowest photometric scatter. We found that for all eight AORs, Aperture photometry gave us the lowest scatter out of the three methods.

%Next, we had to find which aperture radius and modes to use for all the AORs. Indeed, Aperture photometry relies on three hyper-parameters, for which SPCA chooses the best values. Since we are applying photometry one AOR at a time, we need to be consistent in our choice of hyper-parameters. If we let the hyper-parameters vary, we need to account for them with offset hyper-parameters later on, when fitting a model to our data. 
%We found that More details on the photometry methods and hyper-parameters can be found in \cite{belldang2020} and \cite{dang2018}.

\subsection{Astrophysical Model}\label{astro_mod}
Our astrophysical model has four components: a constant stellar flux from the host star, a transit, a secondary eclipse, and a first-order sinusoidal phase variation (see section 4.1 in \citealt{belldang2020} for further details). %, both modelled using the batman package \citep{kreidberg2015} 
We choose to fix astrophysical parameters that are poorly constrained by the Spitzer data to the most up to date values in the NASA Exoplanet Archive\footnote{\url{https://exoplanetarchive.ipac.caltech.edu}}, which can be found in Table \ref{tab:params}: inclination $i$, orbital period $P$, transit midpoint $T_0$, semi-major axis $a$, eccentricity $e$, and argument of periastron $\omega$ \citep{NEA_confirmed}. These values correspond to the ones reported by \cite{Bourrier2018}. We fix the limb-darkening parameters $q_1$ and $q_2$ to \cite{Claret2011} estimates, as used in D16b. We let the following astrophysical parameters of interest freely vary: planet-to-star radius ratio $R_P/R_*$, eclipse depth $F_P/F_*$ and white noise amplitude $\sigma_F$, on which we placed uniform priors from 0 to 1, and the two phase variation parameters $A$ and $B$, on which we placed Gaussian priors \citep[further detail concerning these parameters can be found in][]{belldang2020}. For $A$ the Gaussian prior was $\mu = 0.5, \sigma = 0.5$, and for $B$ the Gaussian prior was $\mu = 0.01, \sigma = 0.5$. 
We experiment with and without constraints on the brightness map of the planet \citep{Keating2017}. More specifically, if we think that the phase curve is not solely planetary in nature, then we do not need to impose a strictly positive phase curve. These unconstrained phase variations, shown in Figure \ref{fig:key3and4}, could be caused by star-planet interactions.

\subsection{Detector Model}
SPCA offers various detector models, but the one we are most interested in is Bi-Linearly Interpolated Subpixel Sensitivity (BLISS) mapping \citep{Stevenson2012}. Firstly, BLISS is a non-parametric detector model, which reduces the computational cost of model fitting. Secondly, BLISS is particularly well-suited for data sets with large spread in the target's centroid, which is our case. Lastly, using other detector models such as polynomial detrending would require us to associate one polynomial to each AOR due to the centroids being so spread out. As a result, the number of parameters in our analysis would drastically increase and so would the computational cost. In comparison, BLISS mapping only has 2 hyper-parameters making it the easiest approach to remove detector systematics from this particular data set. D16b also used BLISS mapping on these data.

The BLISS method maps the detector by associating each centroid to a knot. Choosing the right number of knots is key to our analysis. Too few knots would make our model overly simple and unreliable, while too many knots would lead to over-fitting. We decided to use the same number of BLISS knots as D16b, since having similar detector model parameters ensures that our results are comparable. D16b do not report their number of knots, but they specify that at least 5 centroid position points were linked to each knot. After experimenting with various BLISS knot combinations ($n_x \times n_y$: $21\times16$, $42\times32$, $84\times64$), we determined that $n_x=84, n_y=64$ gave us approximately 5 centroid position points per knot. As noted by D16b, we also find that combining our BLISS mapping approach with a linear function of the width of the PRF along the $x$ and $y$ axes further reduces the level of correlated noise in the photometric residuals.

Finally, we fit our data and estimate the best fit parameters and uncertainties using \texttt{emcee}, a Markov Chain Monte Carlo \citep[MCMC;][]{ForemanMackey2013}. We used 150 MCMC walkers and ran a total of 750,000 burn-in steps until convergence upon visual inspection. We then ran a total of 150,000 production steps to sample the posterior distributions for our final parameter and uncertainty estimates. %A more detailed explanation of the components of each model, and on the choice of hyper-parameters can be found in \cite{dang2018} and \cite{belldang2020}.

\begin{deluxetable*}{cccc}
\tabletypesize{\footnotesize}
\tablecolumns{3}
\tablewidth{0pt}
\tablecaption{Fixed, Free and Derived Parameters for 55 Cancri e \label{tab:params}}
\tablehead{
\colhead{Parameters} & \colhead{SPCA Values} & \colhead{D16b Values}
}
\startdata
\multicolumn{3}{c}{Fixed Astrophysical Parameters} \\ 
$i$ (degrees) & $83.59^{+0.5}_{-0.4}$ & $83.3^{+0.9}_{-0.8}$\\
$P$ (days) & $0.7365474^{+0.0000013}_{-0.0000014}$ & $0.736539 \pm 0.000007$\\
$T_0$ (days) & $2457063.2096^{+0.0006}_{-0.0004}$ & $2455733.013 \pm 0.007$\\
$a/R_*$ & $3.52 \pm 0.01$ & $3.514 \pm 0.62$\\
$e$ & $0.05 \pm 0.03$ & $0.061^{+0.065}_{-0.043}$\\
$\omega$ (degrees) & $86^{+31}_{-33}$ & $202^{+88}_{-70}$\\
$q_1$ & $0.0286$ &  $0.0286$\\
$q_2$ & $0.0554$ & $0.0554$\\
Stellar Effective Temperature (K) & $5172 \pm 18$ & $5250^{+123}_{-172}$\\
Stellar Surface Gravity ($log_{10}(cm/s^2)$) & $4.43 \pm 0.02$ & $4.43^{+0.052}_{-0.14}$\\
Stellar Metallicity (dex) & $0.35 \pm 0.10$ [Fe/H] & 
$0.35 \pm 0.10$ [M/H] \\
checkPhase & True & True\\ \hline
\multicolumn{3}{c}{Free Astrophysical Parameters} \\ 
$R_P/R_*$ & $0.01708^{+0.0016}_{-0.0017}$ & $0.0187 \pm 0.0007$\\
$F_P/F_*$ & $0.000209^{+0.000050}_{-0.000047}$ & $0.000154 \pm 0.000023$\\
Photometric precision (ppm) & $445.4^{+7.5}_{-7.3}$ & $363$\\
A & $0.493^{+0.04}_{-0.07}$ &  Unknown\\
B & $0.108 \pm 0.18$ & Unknown\\ \hline
\multicolumn{3}{c}{Detector Hyper Parameters} \\ 
$x$ knot resolution & $84$ & Unknown\\
$y$ knot resolution & $64$ & Unknown\\ \hline
\multicolumn{3}{c}{Derived Parameters} \\ 
Phase Semi-Amplitude (ppm) &  $110.9^{+17}_{-16}$ & $75.8 \pm 17$\\
Phase Offset (degrees east) & $-12.43^{+21}_{-18}$ & $41 \pm 12$\\
Average Dayside Temperature (K) & $3771^{+669}_{-520}$ & $2999^{+188}_{-193}$\\
Average Nightside Temperature (K) & $1045^{+302}_{-243}$ & $1380 \pm 400$ \\
Conservative Nightside Temperature (K) &
($<1649$ $2\sigma$, $<1951$ $3\sigma$) & --
\enddata
\tablecomments{D16b report phase semi-amplitude and dayside brightness temperature inconsistent with their plotted best-fit model. We re-calculated their semi-amplitude by removing the transit and eclipse from their Source data (see Footnote 4), and measuring the maximum and minimum values. Their dayside brightness temperature was re-calculated using the formula found in \cite{Schwartz2017}. Rather than approximating the spectrum of 55 Cnc as a black body, we used the stellar effective temperature, surface gravity and metallicity reported above to estimate its spectrum using a Phoenix model\footnote{\url{https://www.stsci.edu/hst/instrumentation/reference-data-for-calibration-and-tools/astronomical-catalogs/phoenix-models-available-in-synphot}}. The error bars on the fixed parameters were not used in our analysis. We inflated the error bars on the retrieved parameters for our fiducial analysis according to a binned residuals---Allan---plot (see Figure \ref{fig:allan_plot}). The method used to inflate our uncertainties is detailed in Appendix \ref{sec:appendix_c}.}
\end{deluxetable*}
\section{Results \& Discussion}\label{D&C}
\subsection{Results}\label{res}
\begin{figure*}
    \centering
    \hspace*{-2.8cm}\includegraphics[width=1.3\textwidth]{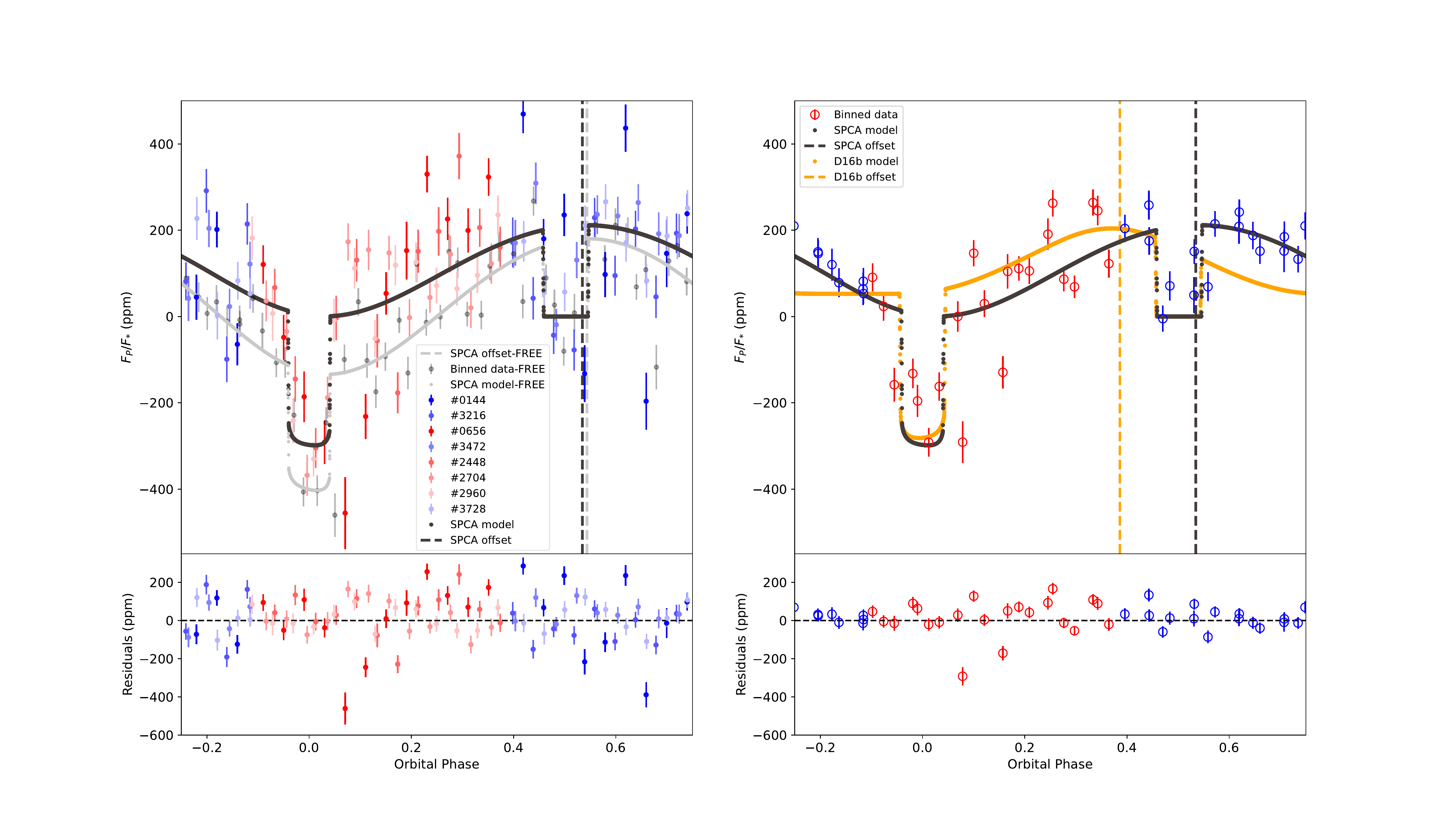}
    \caption{\textbf{Spitzer/IRAC 4.5~$\micron$ phase curve of 55 Cancri e | } \emph{Top left:} Phase-folded, instrument-detrended Spitzer data and our best fit models. The solid grey line corresponds to our unconstrained analysis where we do not assume that the phase curve is planetary in nature and therefore do not impose a strictly positive phase curve. In our legend, we use the abbreviation FREE for this unconstrained analysis. The red and blue points correspond to decorrelated photometric data for all AORs and, like in Figure \ref{fig:key1and2}, we separate them into two groups and distinguish them with different colours. The grey points are the decorrelated photometric data for our unconstrained analysis. For the constrained and unconstrained analyses, the decorrelated photometric data were binned within each AOR by an extra factor of 75 and 150, respectively. \emph{Bottom left:} The residuals for our fiducial, constrained model. \emph{Top right:} The solid black line is our best-fit model using BLISS mapping, same as in the top left plot, and the solid gold line is the best-fit model from D16b. The points represent the decorrelated photometric data for all 8 AORs and, like in Figure \ref{fig:key1and2}, we separate them into two groups and distinguish them with different colours. These decorrelated photometric data were binned across all AORs by an extra factor of 165. \emph{Bottom right:} Corresponding residuals for our best-fit model. The error-bars in each plot are the error on the mean within each bin. We have also added vertical lines for the phase offset obtained with our constrained and unconstrained analyses, and by D16b. Neither our constrained or unconstrained fit match that of D16b, despite using the same raw data detrended with BLISS mapping. The constrained fit is our preferred model; it exhibits a large dayside flux, a negligible offset, and small nightside flux.}
    \label{fig:key3and4}
\end{figure*}

We present the best-fit decorrelated photometry and astrophysical model for the phase variations of 55 Cnc e from our positive-phase-curve fit, shown in the solid black line in Figure \ref{fig:key3and4}. %Our best-fit decorrelated photometry and astrophysical model for the phase variations of 55 Cnc e are presented in the solid black line on Figure \ref{fig:key3and4}. 
We also show D16b's best-fit astrophysical model for comparison. The peak of the phase variation of both resulting model fit are indicated by vertical dashed lines. %We plotted and color-coded our binned photometric data so that it fits what was shown in Figure \ref{fig:key1and2}. 
We list the astrophysical parameter estimates from the MCMC in Table \ref{tab:params}. Our fit shows a transit depth of $291 \pm 56$ ppm, a secondary eclipse depth of $209^{+50}_{-47}$ ppm, a phase semi-amplitude of $111^{+17}_{-16}$ ppm, and a phase offset of $12^{+21}_{-18}$ degrees west. From our resulting model fit, we derived an average dayside temperature of $3,771^{+670}_{-520}$ K and a nightside temperature of $T_{\rm night} < 1649$ K at $2\sigma$. %average nightside temperature of $1,045^{+300}_{-240}$ K. 
Since our unconstrained analysis favors a phase variation with negative flux near transit, the lower uncertainty on the nightside temperature from our constrained analysis is likely underestimated. Therefore, we report a conservative upper limit on the average nightside temperature rather than an estimate. %(further explained in Appendix \ref{sec:appendix_b})}.  

%while the derived parameter values were directly calculated by SPCA after the de-trending process. We note that the values for the phase variation parameters, A and B, as well as the number of knots used for the BLISS Mapping approach, were not specified in the \cite{Demory2016a} analysis, which is why they do not appear in Table \ref{tab:params}.
%Most recently, an analysis of the CHEOPS photometry by \cite{Morris2021} yielded a planet to star radius ratio of $0.01693 \pm 0.00056$, much smaller than the value reported by D16b.
Our best-fit values for the free parameters differ, sometimes significantly, from those reported by D16b, even though we conservatively inflate our uncertainties according to an Allan plot (see comparison in Table \ref{tab:params}). Notably, we conclude that there is no significant offset present in the phase variations. Furthermore, our phase offset does not coincide with the AOR separation at 0.4 orbital phase that can be seen in Figure \ref{fig:key1and2}. The potential causes of these discrepancies are detailed in Appendix \ref{sec:appendix}. A different photometry extraction strategy is the most likely culprit: we adopt a uniform photometric aperture for all AORs, unlike D16b. %Other possible sources of discrepancy include the data reduction process, the fixed astrophysical parameters, the centroiding methods, and the astrophysical models used; although most of these were ruled out, as shown in Appendix \ref{sec:appendix}.}
The observed scatter could also be astrophysical in nature, as both \cite{Demory2016a} and \cite{Tamburo2018} have reported eclipse depth variability. Indeed, the astrophysical variability could have been removed by D16b because they used a different photometric aperture for each AOR.

%Comparing our average photometric precision, \cite{Demory2016b} have a lower value than us, indicating that SPCA's reduction process was more efficient than theirs.(?)%not sure about this

\subsection{Interpretation of Results}
\begin{figure*}
    \centering
    \includegraphics[width = 0.9\linewidth]{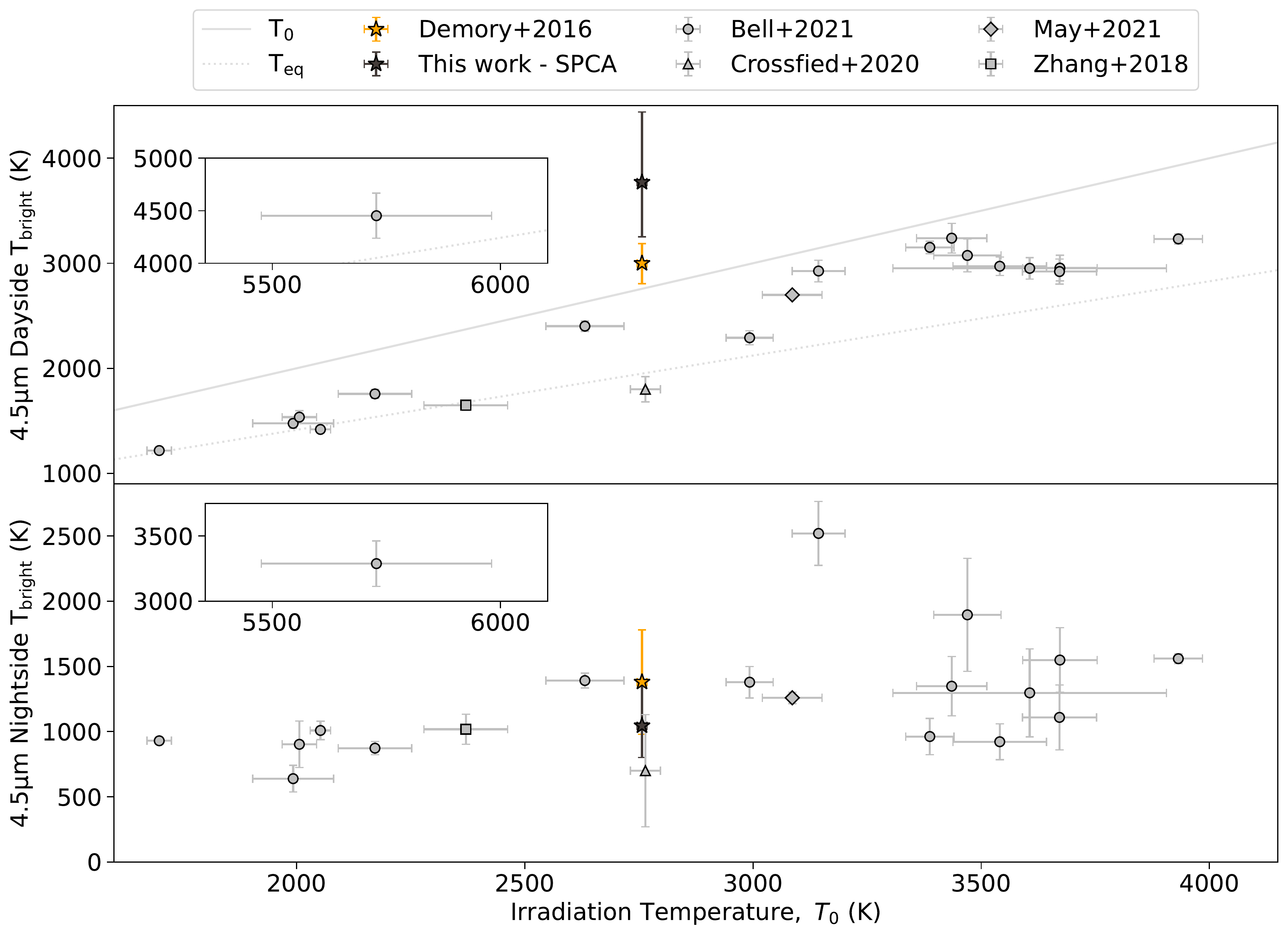}
    \caption{Dayside and nightside brightness temperatures plotted against irradiation temperature, for all circular exoplanets with published 4.5~$\micron$ phase curves. Values plotted as grey circles are from \cite{belldang2020}, while grey triangles, squares and diamonds are from \cite{Crossfield2020}, \cite{Zhang2018} and \cite{May2021}, respectively. The gold and black values for 55 Cnc e are from D16b and this work, respectively. In the top plot, the dotted grey line represents the equilibrium temperature while the solid grey line represents the irradiation temperature. The extremely hot dayside brightness temperature for 55 Cnc e is qualitatively consistent with a temperature inversion and 4.5~$\micron$ emission due to SiO.}
    \label{fig:Taylor}
\end{figure*}

%Binning
%Optical monitoring of 55 Cnc e also enabled the detection of phase modulations. MOST and CHEOPS observations showed modulations with the same period as the planet, but with too large an amplitude to be attributable to the planet \citep{Winn2011, Sulis&Dragomir2019, Morris2021}. %\cite{KippingJansen2020} reported a tentative deep eclipse in TESS data, while \cite{Morris2021} detect the large phase variations---but not the secondary eclipse---in CHEOPS data. 
We adopt the constrained analysis as our fiducial analysis, hypothesizing---as did D16b---that the observed phase variations are solely planetary in nature. 
 Optical monitoring of 55 Cancri e with CHEOPS \citep{Morris2021} and MOST \citep{Winn2011,Sulis&Dragomir2019} revealed phase modulations with `negative' flux values; the system flux drops below the in-eclipse level, which means that the stellar flux is variable. We obtain negative phase variations with our unconstrained analysis, but the semi-amplitude of 172 ppm is greater than that reported in the optical either before \citep[84 ppm;][]{Winn2011} or after \citep[36 ppm;][]{Morris2021} the D16b \emph{Spitzer} observations (Table \ref{tab:params-FREE}). Since it is hard to imagine how stellar variability could impact the IR so much more than the optical, it is unlikely that our unconstrained analysis is affected by the same star-planet interactions as the MOST and CHEOPS observations. For completeness, we report our unconstrained analysis and the best-fit results in Appendix \ref{sec:appendix_b}.

Unlike D16b, our excellent transit depth agreement with \cite{Morris2021} and \cite{MeierValdes2022} tentatively excludes a cloud-free hydrogen-dominated atmosphere (see details in Appendix \ref{sec:Appendix_Transit}). This is consistent with unsuccessful searches for atoms and molecules in transit \citep{Ehrenreich2012,Ridden-Harper2016,Esteves2017,Tabernero2020,Zhang2021,Keles2022}. While \cite{Tsiaras2016} reported a hint of HCN based on Hubble observations, no molecules---including HCN---were detected in ground-based high-dispersion spectroscopy \citep{Deibert2021}. In the balance, existing transit spectroscopy and broadband photometry suggest that 55 Cnc e has either a high mean molecular mass atmosphere, or no atmosphere at all. 

We hypothesize that 55 Cnc e either has a global atmosphere covering both hemispheres of the planet \citep{HammondPierrehumbert2017} or a local dayside atmosphere \citep{leger2011extreme, castan2011atmospheres, kite2016atmosphere, Zilinskas2022}. In either case, the short radiative timescale on 55 Cnc e ensures there is minimal heat transport, which explains the negligible phase offset and nightside flux. The surprisingly high dayside brightness temperature can be explained with the presence of SiO in the atmosphere. Due to 55 Cnc e's close proximity to its host star, silicate species present in a dayside magma ocean evaporate into the atmosphere. Lava planets with silicate atmospheres are prone to temperature inversions due to UV absorption by SiO vapour \citep{ito2015theoretical,Zilinskas2022}. In particular, for a stellar brightness temperature of $5172$ K, the expected dayside brightness temperature at 4.5~$\micron$ is $\sim 3100$ K due to an SiO band, which is consistent with our results. Moreover, \cite{Nguyen2022} showed that the balance between UV heating and cooling of SiO leads to very hot stratospheric temperatures everywhere on the dayside, further enhancing the eclipse depth. 
\section{Conclusion}
We have re-reduced and re-analyzed the 55 Cnc e photometric data collected by Spitzer in 2013 and published by \cite{Demory2016a}. Even though our re-analysis largely followed D16b, our best-fit results, shown in Figure \ref{fig:key3and4}, differ from theirs. We have highlighted the potential origins of these discrepancies in Appendix \ref{sec:appendix}: photometric extraction strategy is the most likely culprit. We found a phase offset of $12^{+21}_{-18}$ degrees west, as opposed to D16b who reported a significant eastward offset, $2\sigma$ away from SPCA's value.  
The negligible phase offset and low nightside flux are consistent with poor heat transport on an ultra-hot planet, while the large dayside flux could be a sign of a dayside inversion due to the SiO absorption of UV and re-emission at 4.5~$\micron$. 

The James Webb Space Telescope (JWST) will soon observe several lava planets including K2-141b \citep{DangProposal2021, EspinozaProposal2021}, GJ 367b \citep{ZhangProposal2021} and 55 Cnc e \citep{HuProposal2021, BrandekerProposal2021}, hopefully answering some questions regarding these exotic worlds.

\section{Acknowledgements}
S.M. acknowledges support from the Rubin Gruber Summer Undergraduate Research Award (SURA). This work would not have been possible without the help of the members of the McGill Exoplanet Characterization Alliance (MEChA), the Institute for Research on Exoplanets (iREx), and the McGill Space Institute (MSI).
L.D. acknowledges support in part through the Technologies for Exo-Planetary Science (TEPS) PhD Fellowship, and the Natural Sciences and Engineering Research Council of Canada (NSERC)'s Postgraduate Scholarships-Doctoral Fellowship. This work is based on archival data obtained with the Spitzer Space Telescope, which is operated by the Jet Propulsion Laboratory, California Institute of Technology under a contract with NASA.
%We would like to thank Lisa Dang and Taylor Bell, without whom we would not have been able to write this paper, and who consistently helped us debug and find a path when we were lost. We'd also like to thank Prof. Nicolas Cowan for his constant encouragement and contagious enthusiasm for this work. 

\facilities{Spitzer Space Telescope}
\software{SPCA, \texttt{batman\footnote{\url{http://lkreidberg.github.io/batman/docs/html/index.html}}} \citep{kreidberg2015}, \texttt{emcee\footnote{\url{https://emcee.readthedocs.io/en/stable/}}} \citep{ForemanMackey2013}, \texttt{corner\footnote{\url{https://corner.readthedocs.io/en/latest/}}} \citep{corner}, Spitzer Heritage Archive \citep{Spitzer_doi}, \texttt{astropy\footnote{\url{https://www.astropy.org}}} \citep{astropy_2013, astropy_2018}, \texttt{numpy\footnote{\url{https://numpy.org}}}} \citep{numpy_2020}

\appendix
\section{Potential Causes for Different Results}\label{sec:appendix}
There are many possible sources for the discrepancies between our best-fit values and those of D16b. As can be seen in Table \ref{tab:params}, we used the most up to date planetary parameters from \cite{Bourrier2018} for our fixed astrophysical parameters. As a test, we also performed an analysis using the same fixed planetary parameters as D16b and found very similar results to the ones shown in Figure \ref{fig:key3and4} and Table \ref{tab:params}, so the fixed planetary parameters are not the dominant source of discrepancy. 
%\begin{itemize}
%    \item \sout{Compare our phase curve parameters to Demory's}
%    \item \sout{compare our analysis to their (Allen plots, ppm precision)}
%    \item \sout{Maybe mention that we used a histogram to find the number of BLISS knots that Demory used.}
%    \item \sout{give a brief explanation on how I plotted the Demory model on top of mine ?}
%    \item \sout{Binning: we do it, maybe they didn't?}
%\end{itemize}

\subsection{Data Reduction}\label{A1}
After the reduction of the Spitzer/IRAC data, D16b grouped the frames in 30 s bins, or by a factor of 1500. As mentioned in Section \ref{Reduce}, we bin the frames by datacube, or by a factor of 64, because it averages over any residual systematic flux modulations present in the datacube \citep{Deming2011}. For computing efficiency we further binned the data by an order of magnitude, bringing our final binning to a factor of 640, or 12.8 s bins. Further details concerning the binning in SPCA, and its effects or lack thereof on detector detrending, can be found in \cite{dang2018} and \cite{belldang2020}.
We ran fits on data that was less binned than D16b, but only by a factor of 2-3 and found similar results as our fiducial analysis. We conclude that binning is unlikely to be the source of discrepancy. 

\subsection{Centroiding}
IRAC detrending depends heavily on centroiding. As a result, choosing different centroiding methods will affect the decorrelation process and impacts the best-fit results. After experimenting with various centroiding methods, including 2D Gaussians, we elected to use a flux-weighted mean as it gave us the lowest photometric scatter. In contrast, D16b used two-dimensional Gaussian fitting, and do not mention testing alternative centroiding schemes. It is possible that our different centroiding strategies led to different astrophysical parameters. 
\subsection{Photometry}
As explained in Section \ref{Photo}, we used a moving, soft-edge aperture of radius 2.2 pixels for our photometric scheme. In comparison, D16b used a fixed aperture with variable radius ranging from 2.6 to 3.4 pixels for each AOR. The aperture radius values for each AOR can be found in the Extended Table 1 of their article. D16b must have accounted for their variable aperture radius by adding an offset parameter for each AOR. It seems likely to us that these additional parameters could explain why we cannot reproduce the D16b results.

\subsection{Astrophysical Model}
D16b use a Lambertian function for the phase modulations of the planet, while SPCA uses a first-order sinusoidal phase variation model. The Lambertian function is an adequate approximation for reflected light, but the Spitzer/IRAC observations correspond to thermal emission. Nonetheless, D16b also experimented with a first-order sinusoidal model and found best-fit results in agreement with their fiducial analysis. All other aspects of the astrophysical model were the same for both analyses.

\section{Unconstrained Analysis}\label{sec:appendix_b}
As optical monitoring with MOST and CHEOPS exhibit phase modulation with amplitude greater than the secondary eclipse depth, we experiment with an unconstrained analysis by allowing the photometric phase variation to drop below the bottom of the eclipse. As explained in section 3.2, the large phase amplitude of our unconstrained analysis, listed in Table \ref{tab:params-FREE} is unlikely to be of the same nature as the phase variation observed in optical observations. %\textbf{Our unconstrained analysis displays a negative nightside temperature (and large uncertainty?) suggesting that our constrained analysis does not probe all values of the phase variations' parameter space. Consequently, we deem our uncertainty on $T_{\rm night}$ too low and we report an upper limit on the nightside temperature.}

\begin{deluxetable}{ccc}[hpbt!]
\tabletypesize{\footnotesize}
\tablecolumns{3}
\tablewidth{0pt}
\tablecaption{Fixed, Free and Derived Parameters for the unconstrained analysis of 55 Cancri e \label{tab:params-FREE}}
\tablehead{
\colhead{Parameters} & \colhead{SPCA Values - unconstrained} & \colhead{D16b Values}
}
\startdata
\multicolumn{3}{c}{Fixed Astrophysical Parameters} \\ 
$i$ (degrees) & $83.59^{+0.5}_{-0.4}$ & $83.3^{+0.9}_{-0.8}$\\
$P$ (days) & $0.7365474^{+0.0000013}_{-0.0000014}$ & $0.736539 \pm 0.000007$\\
$T_0$ (days) & $2457063.2096^{+0.0006}_{-0.0004}$ & $2455733.013 \pm 0.007$\\
$a/R_*$ & $3.52 \pm 0.01$ & $3.514 \pm 0.62$\\
$e$ & $0.05 \pm 0.03$ & $0.061^{+0.065}_{-0.043}$\\
$\omega$ (degrees) & $86^{+31}_{-33}$ & $202^{+88}_{-70}$\\
$q_1$ & $0.028609764174322733$ &  $0.028609764174322733$\\
$q_2$ & $0.05544212401093798$ & $0.05544212401093798$\\
Stellar Effective Temperature (K) & $5172 \pm 18$ & $5250^{+123}_{-172}$\\
Stellar Surface Gravity ($log_{10}(cm/s^2)$) & $4.43 \pm 0.02$ & $4.43^{+0.052}_{-0.14}$\\
Stellar Metallicity (dex) & $0.35 \pm 0.10$ [Fe/H] & $0.35 \pm 0.10$ [M/H] \\
checkPhase & False & True\\ \hline
\multicolumn{3}{c}{Free Astrophysical Parameters} \\ 
$R_P/R_*$ & $0.01623^{+0.00083}_{-0.00084}$ & $0.0187 \pm 0.0007$\\
$F_P/F_*$ & $0.000176^{+0.000025}_{-0.000025}$ & $0.000154 \pm 0.000023$\\
Photometric precision (ppm) & $443.1^{+3.7}_{-3.6}$ & $363$\\
A  & $0.868^{+0.19}_{-0.15}$ & Unknown\\
B  & $0.212^{+0.14}_{-0.13}$ & Unknown\\ \hline
\multicolumn{3}{c}{Detector Hyper Parameters} \\ 
$x$ knot resolution & $84$ & Unknown\\
$y$ knot resolution & $64$ & Unknown\\ \hline
\multicolumn{3}{c}{Derived Parameters} \\ 
Phase Semi-Amplitude (ppm) & $172.2^{+34}_{-28}$ & $75.8 \pm 17$\\
Phase Offset (degrees east) & $-15.68^{+8}_{-7}$ & $41 \pm 12$\\
Average Dayside Temperature (K) & $3439^{+456}_{-397}$ & $2999^{+188}_{-193}$\\
Average Nightside Temperature (K) & $-1185^{+427}_{-485}$ & $1380 \pm 400$%(double check this value)\\
\enddata
\end{deluxetable}

\section{Inflating Uncertainties}\label{sec:appendix_c}
\begin{figure}[hpbt!]
    \centering
    \includegraphics[width = 0.5\linewidth]{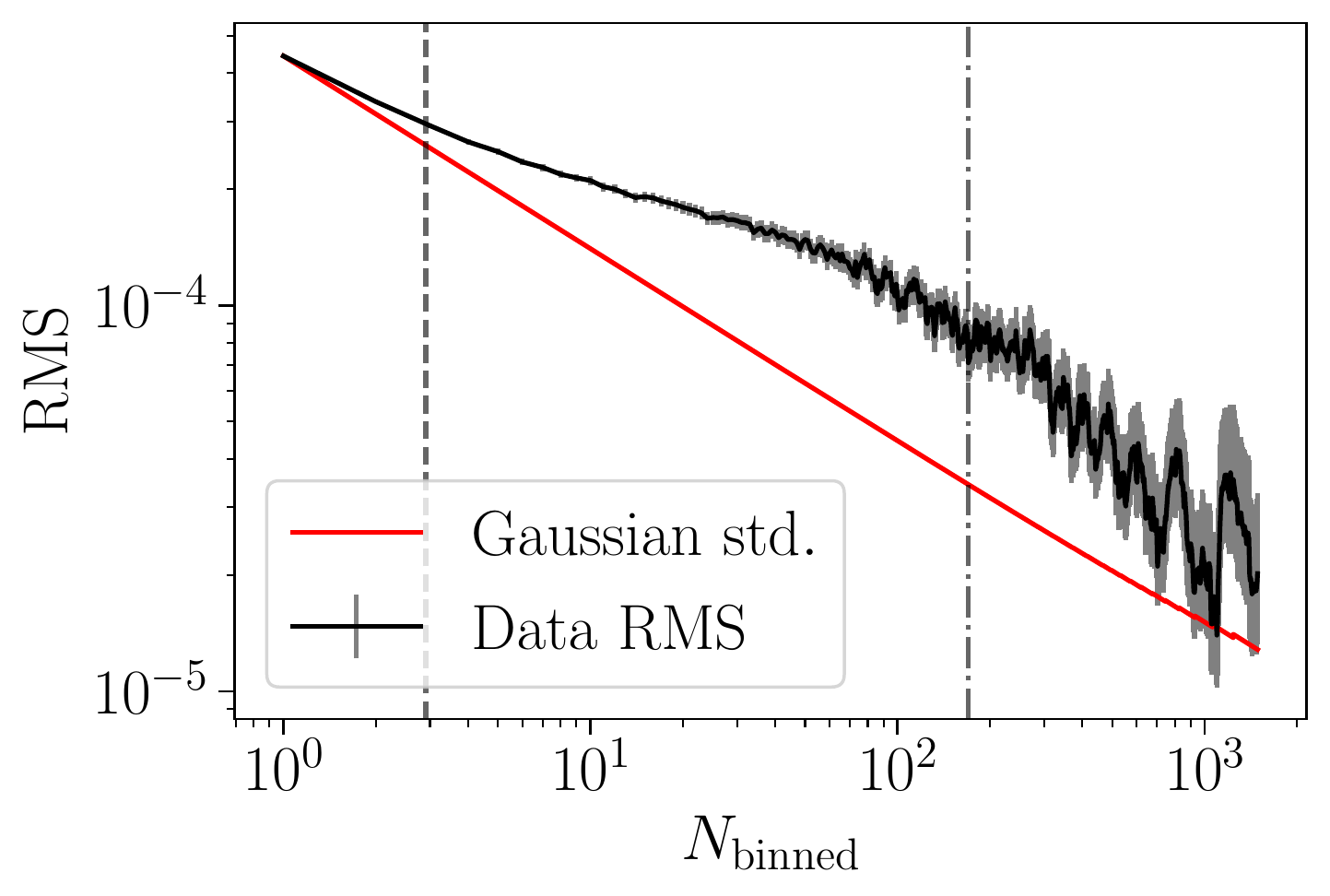}
    \caption{Binned residuals RMS plotted against bin size. The expected decrease in white noise is shown as a solid red line. The solid black line indicates the photometric residual RMS computed for different bin sizes $N_{\rm binned}$. The lighter shaded area around the photometric residual RMS is the uncertainty that the MC3 package computed \citep{Cubillos2017}. The vertical grey dashed and dashdotted lines represent the number of bins contained in the duration of an occultation ingress/egress and in the duration of the entire transit/eclipse.}
    \label{fig:allan_plot}
\end{figure}

We find that there is significant red noise in our residuals and as a result, we choose to inflate our uncertainties following \cite{Pont2006}. The photometric RMS shown in Figure \ref{fig:allan_plot} is about twice as important as the expected Gaussian noise at the vertical dash-dotted line, as such we inflate our error bars by a factor of 2. We inflate the error bars by a factor of 2 for the free astrophysical parameters. For derived parameters, we double the standard deviation of the probability distributions used in the derivation. 

\section{Tentatively Ruling Out Hydrogen-Dominated Atmosphere} \label{sec:Appendix_Transit}
Our planet-to-stellar radius ratio $(R_p/R_*)_{\rm spit} = 0.01708 \pm 0.00168$ is in excellent agreement with the CHEOPS and TESS measurement of $(R_p/R_*)_{\rm cheops} = 0.01693 \pm 0.00035$, and $(R_p/R_*)_{\rm tess} = 0.01708 \pm 0.00024$, respectively \citep{Morris2021, MeierValdes2022}. This suggests that 55 Cnc e either has a high mean molecular mass atmosphere, a cloudy atmosphere, or no atmosphere at all. The atmospheric transit spectral feature can be approximated $2R_p N_H H / R_*^2$, where $N_H$ is the number of atmospheric scale heights probed, and $H=k_B T/\mu g$ is the atmospheric scale height of the planet where $T$ is the atmospheric temperature, $\mu$ is the atmospheric mean molecular mass, and $g$ is the surface gravity \citep{Cowan2015}. Adopting the parametrization of \cite{Cowan2011}, $T = T_*(R_*/a)^{1/2} (1-A_B)^{1/4}(2/3-5\epsilon/12)^{1/4}$ and assuming a Bond albedo of $A_B=0.3$ and heat recirculation efficiency of $\epsilon=0.2$, we estimate the atmospheric temperature of the planet as $T=2290$K. Assuming we probe 4 scale heights, for a hydrogen-dominated atmosphere we estimate the amplitude of the transit spectral feature to be $1.61 \times 10^{-4}$. As the spectral feature is $\sim 2.8$ the quadrature sum of the uncertainties, this tentatively rules out a cloud-free hydrogen atmosphere at 2.8$\sigma$.

\begin{deluxetable}{cccc}[hpbt!]
\tabletypesize{\footnotesize}
\tablecolumns{4}
\tablewidth{0pt}
\tablecaption{Measured Transit Depths of 55 Cnc e\label{tab:transit_depth}}
\tablehead{
\colhead{Parameters} & \colhead{Instrument} & \colhead{$R_p/R_*$} & \colhead{Transit Depth $(R_p/R_*)^2$}
}
\startdata
\cite{Demory2016b} & 4.5$\mu$m Spitzer & $0.0187 \pm 0.0007$ & $0.000350 \pm 0.000026$\\
This work & 4.5$\mu$m Spitzer &  $0.01708 \pm 0.00168$ & $0.000292 \pm 0.000057$\\
\cite{Morris2021} & CHEOPS & $0.01693 \pm 0.00035$ & $0.000287 \pm 0.000012$\\
\cite{MeierValdes2022} & TESS & $0.01708 \pm 0.00024$ & $0.000292 \pm 0.000008$\\
\enddata
\end{deluxetable}


\bibliographystyle{aasjournal}
\bibliography{main}

\label{lastpage}
\end{document}